\documentstyle[12pt]{article}
\begin{document}
\baselineskip=18pt
\begin{center}
{\large{\bf Quasi-classical dynamics of interacting Bose
condensates$^\dag$ }}
\end{center}
\vspace{0.5cm}
\baselineskip=12pt
\begin{center}
A. N. Salgueiro$^{(1)}$, M. C. Nemes$^{(2)}$,

\vspace{.3cm}
 
M. D. Sampaio$^{(2)}$ and A. F. R. de Toledo Piza$^{(1)}$.

\vspace{.5cm}

{\it $^{(1)}$Instituto de F\'{\i}sica,
     Universidade de S\~ao Paulo \\
     CP 20516,\ \ 01452-990 S\~ao Paulo, S.P.,\ \ Brazil \\}

\vspace{.5cm}

{\it $^{(2)}$Departamento de F\'{\i}sica, ICEX,
     Universidade Federal de Minas Gerais\\
     C.P. 702,\ \ 30161-970 Belo Horizonte, M.G.,\ \ Brazil}
\end{center}
\baselineskip=15pt
\vspace{0.5cm}
\begin{center}
{\bf Abstract}
\end{center}
\vspace{0.5cm}

The dynamics of the composition of uniform Bose condensates involving
two species capable of reciprocal interconversion is treated in terms
of a collective quasi-spin model. This collective model quickly
reduces to classical form towards the thermodynamic limit. Quantum
solutions are easily obtained numerically short of this limit which
give insight into the dynamically relevant correlation processes.

\vspace{0.5cm}
\begin{center}
{\it September 1, 1998}
\end{center}

\vspace{1cm}
\noindent PACS numbers: 03.75.Fi, 05.30.Jp, 03.65.Sq, 42.50.Fx

\vspace{\fill}

\noindent\makebox[66mm]{\hrulefill}

\footnotesize 

$^{\dag}$Supported in part by Funda\c{c}\~ao de Amparo \`a Pesquisa do
Estado de S\~ao Paulo (FAPESP) and Conselho Nacional de Desenvolvimento
Cient\'{\i}fico e Tecnol\'ogico (CNPq), Brazil.

\normalsize
\newpage
\baselineskip=15pt

\section{Introduction}

After dilute condensates of bosonic atoms were produced and observed
in the laboratory by Cornell \cite{cor1}, considerable interest arose
concerning the dynamics of the more complex system formed by two
coexisting, coupled condensates \cite{cor2, cor3}. In ref. \cite{cor3}
the dynamics of component separation of a magnetically trapped dual
condensate has been studied by making use of the possibility of
adjusting independently the trapping conditions for each of the two
components. In this case the composition of the mixture remains fixed,
and the finite size of the system plays an essential role. Other
situations, in which the coupling allows for interconversion between
the two condensate types, have also been considered from a theoretical
point of view. Here, even under equilibrium conditions for which the
spatial dynamics of the trapped condensate particles is essentially
frozen, one still has to consider the interesting collective dynamics
of the composition of the dual condensate. One of the situations that
has been considered in this connection involves two different internal
states of the atoms, as in refs. \cite{cor2, cor3}. In this case, the
interconversion coupling is provided by laser-induced Raman
transitions between these two states \cite{zoll}. More recently, the
observation of effects of the so called Feshbach resonances in atomic
Bose condensates \cite{kett} led to the consideration of possible
experimental situations involving coupled atomic and molecular
condensates, the latter occurring in the two-atom channel responsible
for the resonance phenomenon \cite{tthk, ttchk}. In this paper we
explore the fact that, by making use of the assumption of frozen
spatial dynamics, the dynamics of condensate composition can in both
cases be treated in terms of simple collective variables which evolve
in an essentially classical regime, quantum numbers being of the order
of the number of atoms involved in the condensate. We can thus derive
classical, canonical equations of motion governing the model
composition dynamics.

\section{Effective dynamics of coupled condensates}

Although from a microscopic point of view the dual trapped condensate
dynamics involves many quite subtle questions of atomic physics, once
these are duly tamed to the point where they can be manipulated in the
laboratory it is possible to encapsulate their effect in a few
dynamical parameters for the purpose of studying the overall behavior
of the condensate.  Theoretically, the two above mentioned cases have
thus been modeled in terms of an effective Hamiltonian density of the
form \cite{zoll, tthk, ttchk}

\begin{equation}
\label{ham}
{\cal{H}}={\cal{H}}_a+{\cal{H}}_b+{\cal{H}}_c
\end{equation}

\noindent where the first two terms describe the individual
condensates, i.e.

\begin{equation}
\label{hamab}
{\cal{H}}_{a,b}=\phi_{a,b}^\ast\left[-\frac{\hbar^2\nabla^2}{2m_{a,b}}
+\epsilon_{a,b}+\frac{\lambda_{a,b}}{2}\mid\phi_{a,b}\mid^2\right]
\phi_{a,b}
\end{equation}

\noindent and the last term contains the coupling of the two
condensates. For the atom-atom $(AA)$ case it is written as

\begin{equation}
\label{hAAcoupl}
{\cal{H}}_{c}^{(AA)}=\lambda\mid\phi_a\mid^2\mid\phi_b\mid^2 + 
\alpha(\phi_a^\ast\phi_b+\phi_b^\ast\phi_a)
\end{equation}

\noindent while for the atom-molecule (Feshbach resonance, $FR$)
case it reads

\begin{equation}
\label{hFRcoupl}
{\cal{H}}_{c}^{(FR)}=\lambda\mid\phi_a\mid^2\mid\phi_b\mid^2 + 
\alpha(\phi_a^{\ast 2}\phi_b+\phi_b^\ast\phi_a^2).
\end{equation}

\noindent The parameters $\epsilon_{a,b}$ represent possibly
different intrinsic energies in the two boson channels. Their elastic
interaction is described in terms of pseudopotential parameters
$\lambda_{a,b}$, $\lambda$ and channel coupling is represented by the
parameter $\alpha$. In the $AA$ case considered by Zoller \cite{zoll}
the latter involves the intensity of the laser responsible for the
Raman transitions, and should therefore be considered as an externally
determined control parameter. In the $FR$ case, on the other hand,
$\alpha$ stands for the coupling to the quasi-bound molecular state
responsible for the resonance and therefore determining its width,
while $\epsilon_b-2\epsilon_a$ represents the ``detuning'' away from
the resonance, controlled externally by means of an applied magnetic
field \cite{kett}.

For sufficiently extended and uniform systems the effective field
operators $\phi_a$, $\phi_b$ can be usefully expanded in a momentum
basis as

\[
\phi_a(\vec{r})=\frac{1}{\sqrt{V}}\sum_{\vec{k}}e^{i\vec{k}\cdot
\vec{r}}a_{\vec{k}},\;\;\;\;\;\;
\phi_b(\vec{r})=\frac{1}{\sqrt{V}}\sum_{\vec{k}}e^{i\vec{k}\cdot
\vec{r}}b_{\vec{k}}, 
\]

\noindent where the nonhermitean mode operators $a_{\vec{k}}$,
$b_{\vec{k}}$ satisfy standard bosonic commutation
relations. Furthermore, the depletion due to correlations being small
in low density systems \cite{depl}, in the condensate regime
essentially all the bosons are in the zero momentum mode, so that the
relevant Hamiltonian density reduces to the simple single-mode form
(for simplicity we omit the zero momentum label in this case)

\[
{\cal{H}}_{a}\rightarrow\frac{\epsilon_a}{V} a^\dagger a
+\frac{\lambda_a}{2V^2}a^\dagger a^\dagger a a,
\]

\noindent a similar expression for ${\cal{H}}_{b}$ and

\begin{equation}
\label{hcaa}
{\cal{H}}_{c}^{(AA)}\rightarrow\frac{\lambda}{V^2}a^\dagger b^\dagger
b a + \frac{\alpha}{V}(a^\dagger b + b^\dagger a),
\end{equation}

\begin{equation}
\label{hcfr}
{\cal{H}}_{c}^{(FR)}\rightarrow\frac{\lambda}{V^2}a^\dagger b^\dagger
b a + \frac{\alpha}{V^{3/2}}(a^\dagger a^\dagger b + b^\dagger a a)
\end{equation}

\noindent for the atom-atom and Feshbach resonance cases
respectively. In both cases the total number of atoms (namely,
$a^\dagger a + b^\dagger b$ in the $AA$ case and $a^\dagger a + 2
b^\dagger b$ in the $FR$ case) is obviously conserved. Exact single
mode coupled condensate solutions can therefore be obtained through
the diagonalization of finite matrices. Furthermore, the single mode
coupled condensate Hamiltonians belong to the special class of the so
called Curie-Weiss models \cite{cw}, for which the mean-field
approximation becomes exact in the thermodynamic limit (number of
particles $\rightarrow\infty$ at constant density). In order to
explore this latter feature, it is convenient to express
${\cal{H}}^{(AA)}$ and ${\cal{H}}^{(FR)}$ in terms of alternate
dynamic variables, which can be conveniently chosen so as to obey
standard SU(2) commutation relations.

\subsection{Two atomic condensates}

In the $AA$ case, the appropriate variables are just the well known
Schwinger realization of the SU(2) algebra in terms of two types of
bosons \cite{Jschw}

\[
J_z=\frac{1}{2}(a^\dagger a-b^\dagger b);\;\;\;\;\;\;J_+=J_-^\dagger
=a^\dagger b.
\]

\noindent In this case the Casimir operator ${\vec{J}}^2$ appears as
$J(J+1)$ with $J=\frac{1}{2}(a^\dagger a +b^\dagger b)$, half the
number of atoms, and the Hamiltonian density can be written as

\begin{eqnarray}
{\cal{H}}^{(AA)}&=&\frac{\epsilon_a + \epsilon_b}{V}J +
\frac{\epsilon_a - \epsilon_b}{V}J_z
+\frac{\lambda_a}{2V^2}(J+J_z)(J+J_z-1)+\frac{\lambda_b}{2V^2}(J-J_z)
(J-J_z-1) \nonumber \\ &&+\frac{\lambda}{V^2}(J^2-J_z^2)
+\frac{\alpha}{V}(J_+ + J_-). \nonumber
\end{eqnarray}

\noindent In order to take the thermodynamic limit it is more
convenient to work with the scaled Hamiltonian $h^{(AA)}\equiv
V{\cal{H}}^{(AA)}/J$ which reads

\begin{eqnarray}
\label{hAA}
h^{(AA)}&=&\epsilon_a+\epsilon_b+(\epsilon_a-\epsilon_b)\frac{J_z}{J}
+\frac{n\lambda_a}{4}\left(1+\frac{J_z}{J}\right)\left(1+
\frac{J_z}{J}-\frac{1}{J}\right) \nonumber \\ \\
&&+\frac{n\lambda_b}{4}\left(1-\frac{J_z}{J}\right)\left(1-
\frac{J_z}{J}-\frac{1}{J}\right)+\frac{n\lambda}{2}\left(1-
\frac{J_z^2}{J^2}\right)+2\alpha\frac{J_x}{J} \nonumber
\end{eqnarray}

\noindent where the atom density $n=2J/V$ has been introduced. The
thermodynamic limit consists now in letting $J\rightarrow\infty$ at
constant $n$. The spectrum of the scaled components $J_i/J$ remains
bounded in the closed interval [-1,+1] and becomes increasingly dense
as $J$ is increased. The thermodynamic limit corresponds therefore to
the classical limit of a dimensionless angular momentum-like algebra
which may be formally characterized as

\begin{equation}
\label{cllim}
\lim_{J\rightarrow\infty}\frac{J}{i}
\left[\frac{J_k}{J},\frac{J_l}{J}\right]\equiv
\{j_k,j_l\}=\epsilon_{klm}j_m\equiv\epsilon_{klm}
\lim_{J\rightarrow\infty}\frac{J_m}{J}
\end{equation}

\noindent where the curly brackets now denote a Poisson bracket. In
the thermodynamic limit, Eq. (\ref{hAA}) yields the (quasi-)classical
expression for the scaled energy

\begin{eqnarray}
\label{hAAcl}
h_{qc}^{(AA)}&=&\epsilon_a+\epsilon_b+(\epsilon_a-\epsilon_b)j_z
+\frac{n\lambda_a}{4}(1+j_z)^2+\frac{n\lambda_b}{4}(1-j_z)^2
\nonumber \\ &&+\frac{n\lambda}{2}(1-j_z^2)+2\alpha j_x.
\end{eqnarray}

\noindent Alternatively, one can express Eq. (\ref{hAAcl}) as a proper
Hamiltonian function, in terms of a pair of canonically conjugate
variables. One such pair which is quite convenient has been previously
obtained using the quantum kinematical scheme developed by Schwinger
in terms of unitary operator bases \cite{schwkin} and the associate
(discrete) Weyl-Wigner transforms \cite{gelip}. After carrying out the
thermodynamic limit in the way just described, one finds that the
(angle) variable which is canonically conjugate to the (``action'')
variable $j_z$ is just the azimuthal angle $\varphi$, so that $j_x$
can be expressed canonically as

\begin{equation}
\label{canjx}
j_x=\sqrt{1-j_z^2}\cos\varphi.
\end{equation}

\noindent Using this fact, one may derive from Eq. (\ref{hAAcl})
equations of motion for the variables $j_z$ and $\varphi$. A technical
point to be observed here is that the dimensionless character of the
``action'' variable $j_z$ corresponds to measuring time in inverse
energy units. In order to introduce an appropriate time scale one may
write, consistently with Eq. (\ref{cllim}),

\begin{equation}
\label{cltime}
\frac{dj_z}{dt}=\lim_{J\rightarrow\infty}\frac{d}{dt}\frac{J_z}{J}=
\lim_{J\rightarrow\infty}\frac{J}{i\hbar}\left[\frac{J_z}{J},
\frac{V{\cal{H}}^{(AA)}}{J}\right]\equiv\{j_z,
\frac{h_{qc}^{(AA)}}{\hbar}\}
\end{equation}

\noindent which gives 

\begin{eqnarray}
\label{eomAA}
\frac{d\varphi}{dt}&=&
\frac{\partial h_{qc}^{(AA)}}{\hbar\partial j_z}=
\frac{\epsilon_a-\epsilon_b}{\hbar}+\frac{n\lambda_a}{2\hbar}
(1+j_z)-\frac{n\lambda_b}{2\hbar}(1-j_z)-\frac{n\lambda}{\hbar}j_z-
\frac{2\alpha j_z}{\hbar\sqrt{1-j_z^2}}\cos{\varphi} \nonumber \\ \\
\frac{d j_z}{dt}&=&-\frac{\partial h_{qc}^{(AA)}}{\hbar\partial
\varphi}=\frac{2\alpha}{\hbar}\sqrt{1-j_z^2}\sin\varphi. \nonumber
\end{eqnarray}

In order to explore further the significance of the canonical variable
$\varphi$ it is useful to derive the second Eq. (\ref{eomAA}) in the
following alternate way. The (zero momentum) field operators $\phi_a$
and $\phi_b$ satisfy the coupled nonlinear equations

\begin{eqnarray}
i\hbar\dot{\phi}_a&=&\left(\epsilon_a+\lambda_a\mid\phi_a\mid^2+
\lambda\mid\phi_b\mid^2\right)\phi_a+\alpha\phi_b \nonumber \\
i\hbar\dot{\phi}_b&=&\left(\epsilon_b+\lambda_b\mid\phi_b\mid^2+
\lambda\mid\phi_a\mid^2\right)\phi_b+\alpha\phi_a. \nonumber 
\end{eqnarray}

\noindent From these one may easily derive the equations for
$\mid\phi_a\mid^2$ and $\mid\phi_b\mid^2$

\begin{eqnarray}
\label{phieqs}
i\hbar\frac{d}{dt}\mid\phi_a\mid^2&=&2i\alpha\;{\rm Im}\,\phi_b
\phi_a^\ast \nonumber \\ \\
i\hbar\frac{d}{dt}\mid\phi_b\mid^2&=&-2i\alpha\;{\rm Im}\,\phi_b
\phi_a^\ast \nonumber
\end{eqnarray}

\noindent which show an explicit dependence on the phase of
$\phi_b\phi_a^\ast$. Writing this object as  $\mid\phi_b\mid
\mid\phi_a\mid e^{i\delta^{(AA)}}$ the difference of these equations
appears as

\[
\hbar\frac{d}{dt}\left(\mid\phi_a\mid^2-\mid\phi_b\mid^2\right)=
4\alpha\mid\phi_b\mid\mid\phi_a\mid\sin\delta^{(AA)}
\]

\noindent while their sum merely gives the conservation of the total
number of atoms. Using the definitions of $J$ and $J_z$ one has

\[
\mid\phi_a\mid^2-\mid\phi_b\mid^2\equiv\frac{2J_z}{J};\;\;\;\;\;\;
\mid\phi_b\mid\mid\phi_a\mid\equiv\frac{1}{V}\sqrt{(J+J_z)(J-J_z)}
\]

\noindent so that the difference equation becomes

\[
\frac{d}{dt}\frac{J_z}{J}=\frac{2\alpha}{\hbar}\sqrt{1-
\frac{J_z^2}{J^2}}\sin\delta^{(AA)}.
\]  

\noindent This corresponds to the second Eq. (\ref{eomAA}) and
furthermore identifies the angle variable $\varphi$ with the phase
$\delta^{(AA)}$.

The density dependence of the Hamiltonian Eq. (\ref{hAAcl}) also
gives a direct analytical expression for the pressure exerted by the
interacting condensates. One has in fact

\begin{equation}
\label{pAA}
P^{(AA)}\equiv n^2\frac{\partial h_{qc}^{(AA)}}{\partial n}=
\frac{n^2}{4}\left[\lambda_a(1+j_z)^2+\lambda_b(1-j_z)^2+2
\lambda(1-j_z^2)\right]
\end{equation}

\noindent In order to avoid the collapse of stationary states this
expression must be positive when evaluated at the equilibrium value of
$j_z$.

\subsection{Feshbach-resonant atomic-molecular condensates}

The $FR$ case, on the other hand, can be treated in precisely the same
way once an appropriate realization of the SU(2) algebra is
constructed. To this effect, consider the case in which there are $N$
atoms present. The relevant finite dimensional space in which
${\cal{H}}^{(FR)}$ is to be diagonalized is then generated by base
vectors of the form $\mid N-2n_b,\;n_b\rangle$, where the two labels
denote the number of atoms and the number of molecules
respectively. For definiteness, $N$ will be assumed to be even, so that
$n_b$ runs from zero to $N/2$. The next step is to identify this basis
with the $J_z$ eigenstates of a SU(2) multiplet associated with the
eigenvalue $N/4(N/4+1)$ of the Casimir operator. A convenient way of
doing so is to identify the eigenvalue of $J_z$ with $n_b-N/4$, so
that $J_z=\frac{1}{4}(2b^\dagger b-a^\dagger a)$ and

\[
\mid N-2n_b,\;n_b\rangle\leftrightarrow\mid
J=\frac{N}{4},\;J_z=n_b-\frac{N}{4}\rangle.
\]

\noindent If one then defines $J_\pm$ in terms of their standard
action on the $J_z$ eigenstates, i.e.

\[
J_\pm\mid J,J_z\rangle=\sqrt{J(J+1)-J_z(J_z\pm 1)}\mid J,J_z\pm
1\rangle
\]

\noindent one finds, after a straightforward calculation,

\[
J_+=\frac{1}{\sqrt{2(a^\dagger a+1)}}aab^\dagger;\;\;\;\;\;\;
J_-=a^\dagger a^\dagger b\frac{1}{\sqrt{2(a^\dagger a+1)}}.
\]

\noindent The $FR$ Hamiltonian density can now be written in the form

\begin{eqnarray}
{\cal{H}}^{(FR)}&=&\frac{\epsilon_b+2\epsilon_a}{V}J+
\frac{\epsilon_b-2\epsilon_a}{V}J_z+\frac{\lambda_a}{V^2}2(J-J_z)
(J-J_z-1)+\frac{\lambda_b}{2V^2}(J+J_z)(J+J_z-1) \nonumber \\
&&+\frac{\lambda}{V^2}
2(J^2-J_z^2)+\frac{\alpha}{V^{3/2}}\left(J_-\sqrt{4(J-J_z)+2}+
\sqrt{4(J-J_z)+2}\;J_+\right) \nonumber
\end{eqnarray}

\noindent and the corresponding $J$-scaled Hamiltonian, after taking
the thermodynamic limit, becomes

\begin{eqnarray}
\label{hFRqc}
h_{qc}^{(FR)}&\equiv&\lim_{J\rightarrow\infty}
\frac{V{\cal{H}}^{(FR)}}{J} \\
&=&\epsilon_b+2\epsilon_a+(\epsilon_b-2\epsilon_a)j_z+
\frac{n\lambda_a}{2}(1-j_z)^2+\frac{n\lambda_b}{8}(1+j_z)^2+
\frac{n\lambda}{2}(1-j_z^2) \nonumber \\
&&+2\alpha\sqrt{n}\sqrt{1-j_z}\;j_x \nonumber
\end{eqnarray}

\noindent where the vanishing of the commutators $[\frac{J_z}{J},\;
\frac{J_\pm}{J}]$ in this limit has been used to obtain the last
term. An interesting feature of these Hamiltonians is the quenching of
the coupling term involving $\alpha$ when the molecular component is
dominant ($J_z$ close to J or $j_z$ close to 1). It can be traced
ultimately to the feature of Bose statistics which associates the
factor $\sqrt{(N-2n_b+1)(N-2n_b+2)n_b}$ to the amplitude for
converting one molecule into two atoms when there are $n_b$ molecules
present. This factor in fact decreases faster than that associated
with the plain $J_-$ operator when $n_b$ is large (close to its
maximum value $N/2$).

One may next adopt again the canonical form of $j_x$ given in
Eq. (\ref{canjx}) and the appropriate time scale to write the
equations of motion

\begin{eqnarray}
\label{eomFR}
\frac{d\varphi}{dt}&=&\frac{\epsilon_b-2\epsilon_a}{\hbar}-
\frac{n\lambda_a}{\hbar}(1-j_z)+\frac{n\lambda_b}{4\hbar}(1+j_z)
-\frac{n\lambda}{\hbar}j_z-\frac{\alpha\sqrt{n}}{\hbar}
\frac{1+3j_z}{\sqrt{1+j_z}}\cos\varphi \nonumber \\ \\
\frac{dj_z}{dt}&=&\frac{2\alpha\sqrt{n}}{\hbar}(1-j_z)
\sqrt{1+j_z}\sin\varphi. \nonumber
\end{eqnarray}

\noindent In a way which is completely analogous to the $AA$ case one
may here identify the canonical variable $\varphi$ with the phase
$\delta^{(FR)}$ defined as

\[
\phi_a^2\phi_b^\ast\equiv\mid\phi_a\mid^2\mid\phi_b\mid\;
e^{i\delta^{(FR)}}.
\]

The pressure can also be obtained in this case with the result

\begin{equation}
\label{pFR}
P^{(FR)}=\frac{n^2}{2}\left[\lambda_a(1-j_z)^2+\frac{\lambda_b}{4}
(1+j_z)^2 +\lambda(1-j_z^2)\right]+\alpha
n^{3/2}(1-j_z)\sqrt{1+j_z}\cos\varphi,
\end{equation} 
 
\noindent which reproduces (with different notation) the result
obtained in ref. \cite{tthk}. When the square brackets are positive
for the equilibrium value of $j_z$, one may still have a domain of
negative pressures at low densities when the coefficient of $n^{3/2}$
is negative. Also as observed in \cite{tthk}, this implies that the
system becomes ``self-bound'' and saturates at the value of $n$ for
which the pressure vanishes.

In order to make the model sufficiently realistic for the $FR$ system
it is important to take into account the loss of atoms which may be
expected due to an enhanced rate of inelastic collisions involving the
molecular channel directly \cite{kett,ttchk}. This loss has in fact been
used as a signal to detect the Feshbach resonance in
ref. \cite{kett}. The simplest way of including the loss of atoms
consists in adding appropriate master loss-terms to the equations of
motion for the atomic and molecular densities, $n_a$ and $n_b$. Under
loss-free conditions, we may use the expressions

\begin{equation}
\label{pardens}
n_a=\frac{1-j_z}{2}n,\;\;\;\;\;\;n_b=\frac{1+j_z}{4}n
\end{equation}

\noindent and the equation of motion for $j_z$, Eq. (\ref{eomFR}), to
write

\begin{eqnarray}
\frac{dn_a}{dt}&=&-\frac{n}{2}\frac{dj_z}{dt}=
-\frac{\alpha n^{3/2}}{\hbar}(1-j_z)\sqrt{1+j_z}
\sin\varphi \nonumber \\
\frac{dn_b}{dt}&=&\frac{n}{4}\frac{dj_z}{dt}=
\frac{\alpha n^{3/2}}{2\hbar}(1-j_z)\sqrt{1+j_z} 
\sin\varphi. \nonumber
\end{eqnarray}

\noindent These equations are the $FR$ analogs of Eqs. (\ref{phieqs}),
used to identify the canonical variable $\varphi$ with the phase of an
appropriate combination of field variables. To take loss effects into
account these equations are replaced by

\begin{eqnarray}
\frac{dn_a}{dt}&=&-\frac{\alpha n^{3/2}}{\hbar}(1-j_z)\sqrt{1+j_z}
\sin\varphi-c_{aa}n_a^2-c_{ab}n_bn_a \nonumber \\
\frac{dn_b}{dt}&=&\frac{\alpha n^{3/2}}{2\hbar}(1-j_z)\sqrt{1+j_z}
\sin\varphi-c_{bb}n_b^2-c_{ba}n_an_b \nonumber 
\end{eqnarray}

\noindent which involve four loss-rate coefficients $c_{ij}$. These
are defined so that $c_{ij}n_j$ represents the decay constant of the
$i$-boson density due to collisions with $j$-bosons. The important
decay rates in the molecular channel thus imply larger values of
$c_{bb}$ and $c_{ba}$ when compared with $c_{aa}$ and $c_{ab}$.  The
time evolution of the total density $n$ is now determined by the
equation

\begin{equation}
\label{master}
\frac{dn}{dt}=2\frac{dn_b}{dt}+\frac{dn_a}{dt} \nonumber \\
=-\frac{n^2}{8}\left[2c_{aa}(1-j_z)^2+c_{bb}(1+j_z)^2+
(c_{ab}+c_{ba})(1-j_z^2)\right]
\end{equation}

\noindent which must be solved together with Eqs. (\ref{eomFR}) in
order to include atomic losses in the dynamics of the interacting
condensates. It should be noted that this procedure corresponds
exactly to that adopted in ref. \cite{ttchk}, where one works with the
nonlinear equations of motion for the fields $\phi_a$, $\phi_b$
modified to have complex pseudopotential parameters. In fact, the
terms involving the imaginary parts of these parameters cancel from
the equation of motion for the variable $j_z$, but give non vanishing
contributions to $dn/dt$ which exactly reproduce Eq. (\ref{master}).

\section{Numerical results}

In order to explore the classical propensities of the quantum,
Curie-Weiss coupled condensate dynamics, it is most convenient to make
use of a phase-space quantum description. An appropriate description
of this sort, for systems evolving in quantum phase spaces of finite
dimensionality, consists of the discrete action-angle Weyl-Wigner
representation used in Ref. \cite{gelip}. In the next few lines we
simply collect the final prescription to obtain the discrete
transforms, and refer the interested reader to this reference for
further details. For a state $\mid a\rangle$ represented in the
appropriate multiplet base $\mid J,m\rangle$ as

\[
\mid a\rangle=\sum_{m=-J}^J a_m\mid J,m\rangle
\]

\noindent one first constructs the matrix

\[
r(k,l)=\frac{1}{\sqrt{2J+1}}\sum_{m=-J}^J a_m a_{\{m+l\}}^\ast
\exp\left[-\frac{2\pi i}{2J+1}k(m+\frac{l}{2})\right]
\]

\noindent where the range of the integers $k$ and $l$ is $-J\leq
k,l\leq J$ and the index $\{m+l\}$ denotes the value of $m+l$
cyclically confined to the range $-J,\;J$ of the basis
labels. Explicitly, one has

\[
\{m+l\}=m+l-(2J+1)\;\;{\rm Floor}\left(\frac{m+l+J}{2J+1}\right)
\]

\noindent where Floor($x$) denotes the {\it larger} integer (negative
for $x<0$) less than or equal to $x$. The desired discrete Wigner
phase-space representative $a_w(p,q)$ of the state $\mid a\rangle$ is
then obtained as the double (discrete) Fourier transform

\[
(2J+1)\,a_w(p,q)=\frac{1}{\sqrt{2J+1}}\sum_{k,l}\exp\left[
\frac{2\pi i}{2J+1}(pk+ql)\right]r(k,l).
\]

\noindent In this expression the range of the integers $p$ and $q$ is
also bounded as $-J\leq p,q\leq J$, and the properly scaled variables
corresponding to $j_z$ and $\varphi$ are $q/J$ and $2\pi p/(2J+1)$
respectively. The Weyl transform of the Hamiltonian can be obtained in
exactly the same way, replacing the amplitude products $a_m
a_{\{m+l\}}^\ast$ by the matrix elements $\langle J,m\mid V{\cal H}/J
\mid J,\{m+l\}\rangle$ multiplied by the number of states $2J+1$, when
evaluating $r(k,l)$.

\subsection{Two atomic condensates}

We restrict our numerical treatment of the composition dynamics of two
atomic condensates to the ``symmetric'' case, in the sense of
ref. \cite{zoll}, i.e. $\epsilon_a=\epsilon_b\equiv 0$ (implying a
suitable definition of the energy scale) and $\lambda_a=\lambda_b
\equiv\lambda_0 $. Furthermore, we adopt units such that the atom
density $n=1$ and $\hbar=\lambda_0=1$ so that energies are given in
units of $\lambda_0 n$, times are given in units of $\hbar/\lambda_0
n$ and the various possible situations will unfold by varying the
remaining parameters, $\lambda$, $\alpha$ and, when short of the
thermodynamic limit, $J\equiv N/2$, cf. Eqs. (\ref{hAA}) and
(\ref{hAAcl}).

The numerical spectrum of $h^{(AA)}$, Eq. (\ref{hAA}), is given for
$\lambda=1.5$, $\alpha=.1$ and $J=20$ in Fig. 1, together with the
mean $b$-type atom numbers $\langle{n}_b\rangle\equiv \langle
b^\dagger b\rangle$ and variances $\sigma_b\equiv\sqrt{\langle
(b^\dagger b)^2\rangle - \langle{n}_b\rangle^2}$ for each of the
corresponding eigenstates. The states with energy $E_k<1$ are in fact
nearly degenerate doublets (see Fig. 5 below) of ``Schr\"odinger cat
states'', consisting of superpositions of basis states with
$\langle{n}_b\rangle\sim 0$ and $\langle{n}_b\rangle \sim N$
\cite{zoll}. The large values of $\sigma_b$ for these states signal
the large, strongly correlated fluctuations of $n_b$ and $n_a$. The
doublet structure can be immediately understood with reference to the
Weyl transform of $h^{(AA)}$, shown in Fig. 2 together with the
quasi-classical energy surface corresponding to $h^{(AA)}_{qc}$: it
results from the symmetric minima near $j_z=\pm 1$ separated by the
barrier which peaks at $j_z=0$. As also discussed in ref. \cite{zoll},
this situation results from having $\lambda>\lambda_0$ {\it and} small
enough $\alpha$ (``weak laser'' case). The energy splitting of the
doublet members approaches zero as $J$ is increased, leading to
degeneracy in the thermodynamic limit. The Wigner functions
corresponding to each state in the lowest doublet for $J=20$ are shown
in Fig. 3. The lowest (highest) member of the doublet involves a
symmetric (anti-symmetric) superposition of states $\mid J,M\rangle$
strongly concentrated on the largest values of $\mid M\mid$. The
Wigner functions show moreover that these states are strongly peaked
also in the conjugate angle variable, as a result of the dips of the
energy surface at $\varphi=\pm \pi$. Increasing $\alpha$ will depress
the pass along the $\varphi=\pi$ line leading eventually to a minimum
at $j_z=0$ and therefore to ground states dominated by small $\mid
M\mid$ components.  This is shown in Fig. 4, where the probabilities
$p(M)\equiv\mid\langle J,M\mid{\rm g.s.}\rangle\mid^2$ are plotted
against $M$ for a range of values of $\alpha$.

A most dramatic consequence of the strong collective character of the
composition dynamics as described by the Hamiltonian $h^{(AA)}$,
Eq. (\ref{hAA}), is that, as was pointed out before (see
e.g. ref. \cite{baires}), a suitably scaled distribution of the energy
eigenvalues quickly approaches the quasi-classical distribution in
which the fraction of levels with eigenvalue smaller than a given
energy is proportional to the area of the phase-space domain in which
the quasi-classical energy is less than this value. This ultimately
allows associating energy eigenstates with phase-space trajectories on
an essentially one to one basis. Fig. 5 illustrates this feature
showing the fraction of the total number of states having energy less
than $E$ as a function of $E$ for several values of $J$ together with
the quasi-classical limit. One sees there that even the case N=20 is
already fairly close to the quasi-classical limit from which it
differs mainly by an overall translation in energy related to the
terms of order $1/J$ in Eq. (\ref{hAA}). Also visible is the
staggering associated with the doublet structure of the
$\lambda>\lambda_0$, ``weak laser'' case, which persists up to
energies close to that corresponding to the separatrix going through
the highest point of the $\varphi=\pi$ pass at $j_z=0$ ($E=1.05$ for
this example, in the quasi-classical limit). An enhancement of the
level density at this energy is also clearly visible.

\subsection{Feshbach-resonant atomic-molecular condensates}

While in the $AA$ case the relevant external control parameter is the
intercondensate coupling $\alpha$, associated with laser intensity, in
the case of realistic $FR$ systems the relevant parameter is the
detuning $\delta\equiv\epsilon_a-2\epsilon_b$, controlled by an
externally applied uniform magnetic field. In order to illustrate the
composition dynamics in this case we take $\lambda_a=\lambda_b=
\lambda=1$ and again choose units such that $\hbar=n=1$, the zero of
the energy scale being defined so that $\epsilon_a=0$. The remaining
parameters to be considered are therefore $\alpha$, here related to
the resonance width, the detuning parameter $\epsilon_b$, which is the
intrinsic energy difference between one molecule and two atoms, and
(again when short of the thermodynamic limit) $J\equiv N/4$.

The numerical quantum spectrum of $V{\cal{H}}^{(FR)}/J$ is given in
Fig. 6 (similar to Fig. 1, corresponding to the $AA$ case) for $\alpha
=\sqrt{2}/2$, $\epsilon_b=2$ and $J=20$. The mean molecule number
$\langle n_b\rangle$ and variance $\sigma_b$ for the eigenstates are
shown at the corresponding energy eigenvalues. The Weyl transform of
the quantum Hamiltonian is shown in Fig. 7, together with its classical
limit, Eq. (\ref{hFRqc}). The scaling property of the level density
with $J$, analogous to that shown in Fig. 5 for the $AA$ case, is
shown in Fig. 8.

In this case, the quantum ground state is strongly localized, both in
action and angle variables, in the non-degenerate minimum of the
energy surface, as shown by its Wigner function, Fig. 9. Increasing
the value of the detuning parameter energetically favors the lower $M$
components in the ground state, which becomes eventually an
essentially pure atomic condensate, as can be seen in Fig. 10, where
$\epsilon_b=50$ with no change in the remaining parameters. This
figure shows the Weyl energy surface for this case and also the Wigner
function of the corresponding ground state. Since this energy surface
is dominated by the detuning term, and therefore rendered only very
weakly dependent on the angle variable, its ground state appears as
strongly delocalized in this variable for $J=20$. The definition of
this ground state in the angle variable is however expected to
increase for larger values of $J$, as a more collective state fits in
the very shallow minimum which persists at $\varphi=\pi$, near
$j_z=-1$.

Next we use the $\epsilon_b=50$ quantum ground state as an initial
condition in order to obtain numerically the (loss free) quantum
evolution under the $\epsilon_b=2$ Hamiltonian. This corresponds to
the scheme devised in ref. \cite{ttchk}, in which the ground state is
prepared at some initial large value of the detuning which, is then
suddenly reduced by adjusting the applied external magnetic field. The
absolute square coefficients of the decomposition of the initial state
in the $\epsilon=2$ eigenstates are shown in Fig. 11, and the
resulting time dependence of the mean value of $J_z/J$ for $J=20$ are
shown in Fig. 12 together with the corresponding time-dependent
variances. The mean number of atoms and of molecules, $\langle
n_a\rangle$ and $\langle n_b\rangle$, are related to $\langle
J_z\rangle/J$ simply as

\[
\langle n_a\rangle=2J\left[1-\frac{\langle J_z\rangle}{J}\right],
\;\;\;\;\;\;\langle n_b\rangle=J\left[1+\frac{\langle J_z\rangle}{J}
\right]
\]

\noindent and will therefore oscillate with opposite phases as a result
of the conservation of the total number of atoms in this
calculation. Also shown in the same plot is the function $j_z(t)$
obtained by integrating numerically Eqs. (\ref{eomFR}) with
$\epsilon_b=2$, $\alpha=\sqrt{2}/2$ and using the phase-space
coordinates of the quasi-classical equilibrium point corresponding to
$\delta=50$ as initial conditions. Note that, in view of
Eq. (\ref{cltime}), the appropriate time parameter to be used in the
quantum calculation in order to compare with the quasi-classical
result corresponds to solving the Heisenberg equation of motion

\[
\frac{i\hbar}{J}\frac{d}{dt}\left[\frac{J_z}{J}\right]=
\left[\frac{J_z}{J}, \frac{V{\cal{H}}^(FR)}{J}\right]
\]

\noindent with $J=20$ in this case. The similarity of the two graphs
illustrates the relevance of the quasi-classical calculation of
collective properties of the system already for $J=20$.

Finally, we include the phenomenological loss terms in the
quasi-classical equations of motion by integrating simultaneously
Eqs. (\ref{eomFR}) and Eq. (\ref{master}) with the realistic values of
the loss-rate coefficients used in ref. \cite{ttchk}, namely $c_{aa}=
c_{ab}=0$ and $c_{bb}=c_{ba}=0.5$, in the units adopted here. Fig. 13
shows the resulting time evolution of $j_z(t)$ and of $n(t)$, the
latter quantity being measured in units of the initial density
$n(t=0)=1$. The atomic and molecular densities are now given by
Eq. (\ref{pardens}). Their decrease is due mainly to the decrease of
the total density $n$ since, as can be seen in Fig. 13, the amplitude
of the oscillations of $j_z$ remains relatively immune to the loss
effects. Furthermore, the undulating behavior in the decrease of
$n(t)$ reflects the asymmetry of the adopted values of the loss-rate
coefficients, as a result of which the losses take place in the
molecular channel only. This in fact causes the effective loss rate of
$n(t)$ to vanish as $j_z(t)$ approaches its lower bound.

\section{Concluding comments}

The fact that the composition dynamics of essentially undepleted,
coupled uniform Bose condensates can be completely reduced to a very
compact form in terms of collective variables only is a direct
consequence of the Curie-Weiss character of the underlying model
Hamiltonians. While the terms included in these Hamiltonians have been
restricted here to those considered earlier \cite{zoll,tthk}, the
inclusion of other terms, such as two molecule collisions leading to
four condensate atoms and atom-molecule collisions leading to three
condensate atoms in the Feshbach resonance case, is completely
straightforward as they can be readily expressed also in terms of the
same collective variables. Not only the resulting collective
Hamiltonians reduce to classical form in the thermodynamic limit, but
it it also an easy matter to obtain complete numerical quantum
solutions short enough of this limit. The study of these solutions
indicates that the quasi-classical thermodynamic limit is approached
rather fast, while allowing for insight into the dynamically relevant
correlation processes.

It should be kept in mind, however, that purely collective dynamical
treatments of this kind involve a severe truncation of the full model
problem as stated e.g. in Eqs. (\ref{ham}) to (\ref{hFRcoupl}). In
particular, collective excitations will in general couple to depletion
effects which fall beyond a purely collective description. We believe
nevertheless that the remarkable simplicity of the latter readily
provides useful information on such systems and may give useful
indications for the development of more complete treatments.

\newpage
\noindent 
\Large{\bf Figure Captions}

\vspace{1cm}
\normalsize
\noindent 
{\bf Figure 1.} Energy spectrum of $h^{(AA)}$, Eq. (\ref{hAA}), for a
symmetric, ``weak-laser'' case. Units and parameter values are
explained in the text. The points below $E\simeq 1$ correspond to
nearly degenerate doublets (cf. Fig. 5). Mean numbers of $b$ atoms
and the respective variances are shown at the corresponding energy
eigenvalues.

\medskip
\noindent 
{\bf Figure 2.} Discrete action-angle Weyl transform of the
Hamiltonian used in Fig. 1, (a), and the corresponding quasi-classical
energy surface, (b).

\medskip
\noindent 
{\bf Figure 3.} Discrete action-angle Wigner functions corresponding
to the lower (a) and upper (b) member of the lowest doublet in Fig. 1.

\medskip
\noindent 
{\bf Figure 4.} Evolution of the probabilities p(M) of the various
$\mid J,M\rangle$ components of the ground-state as a function of the
``laser strength'' control parameter $\alpha$. The double-peaked
distributions correspond to ``Schr\"odinger cat states''. 

\medskip
\noindent 
{\bf Figure 5.} Evolution of the integrated level density as a
function of $J$. Other parameters are as in Fig. 1. The
quasi-classical limit is also shown.

\medskip
\noindent 
{\bf Figure 6.} Typical ``small detuning'' energy spectrum of
$V{\cal{H}}^{(FR)}/J$. Units and parameter values are explained in the
text. Mean numbers of molecules and the respective variances are shown
at the corresponding energy eigenvalues.

\medskip
\noindent 
{\bf Figure 7.} Discrete action-angle Weyl transform of the
Hamiltonian used in Fig. 6, (a), and the corresponding quasi-classical
energy surface, (b). 

\medskip
\noindent 
{\bf Figure 8.} Evolution of the integrated level density as a
function of $J$. Other parameters are as in Fig. 6.

\medskip
\noindent 
{\bf Figure 9.} Discrete action-angle Wigner function corresponding to
the ground state of the spectrum shown in Fig. 6.

\medskip
\noindent 
{\bf Figure 10.} Weyl transform of the Hamiltonian, (a), and Wigner
function of the ground state, (b), for ``large detuning''
$\epsilon_b=50$. Other parameters as in Fig. 6.

\medskip
\noindent 
{\bf Figure 11.} Squared expansion amplitudes of the ``large
detuning'' ground state of Fig. 10 in terms of the ``small detuning''
eigenstates corresponding to Fig. 6. Results are given for the full
$J=20$ spectrum.

\medskip
\noindent 
{\bf Figure 12.} Exact quantum evolution of the ``large detuning''
ground state under the ``small detuning'' Hamiltonian for $J-20$. The
heavy curve shows the time evolution of $\langle j_z\rangle$, and the
error band shows the time evolution of the root-mean-square dispersion
of $j_z$. Also shown (white squares) is the result of an integration
of the quasi-classical Eqs. (\ref{eomFR}). See text for details.

\medskip
\noindent 
{\bf Figure 13.} Numerical solution of the coupled quasi-classical
equations (\ref{eomFR}) and (\ref{master}). See text for details.


\begin{thebibliography}{99}

\bibitem{cor1} M. H. Anderson, J. R. Ensher, M. R. Matthews,
C. E. Wieman and E. A. Cornell, Science {\bf 269}, 198 (1995).

\bibitem{cor2} C. J. Myatt, E. A. Burt, R. W. Ghrist, E. A. Cornell
and C. E. Wieman, Phys. Rev. Lett. {\bf 78}, 586 (1997).

\bibitem{cor3} D. S. Hall, M. R. Matthews, J. R. Ensher, C. E. Wieman
and E. A. Cornell, e-print cond-mat/9804138. 

\bibitem{zoll} J. I. Cirac, M. Lewenstein, K. M{\o}lmer and P. Zoller,
Phys. Rev {\bf A57}, 1208 (1998).

\bibitem{kett} S. Inouye, M. R. Andrews, J. Stenger, H.-J. Miesner,
D. M. Stamper-Kurn and W. Ketterle, Nature {\bf 392}, 151 (1998). 

\bibitem{tthk} P. Tommasini, E. Timmermans, M. S. Hussein and
A. K. Kerman, e-print cond-mat/9804015.

\bibitem{ttchk} E. Timmermans, P. Tommasini, R. C\^ote,
M. S. Hussein and A. K. Kerman, e-print cond-mat/9805323.

\bibitem{depl} Y. Castin, R. Dum,  Phys. Rev. Lett. {\bf 79}, 3553 (1997);
P. Nozi\`eres in ``Bose-Einstein Condensation'´, Cambridge
(1998), ed. A. Griffin, D. W. Snoke and S. Stringari, 15. 

\bibitem{cw} see e.g. L. van Hemmen, Fortschritte der Physik {\bf 26},
397 (1978).

\bibitem{Jschw} J. Schwinger, Quantum Theory of Angular Momentum,
L. Biedenharn and H. Van Dam Eds., Academic Press, New York 1965. 

\bibitem{schwkin} J. Schwinger, Quantum Kinematics and Dynamics,
W. A. Benjamin, Inc. New York 1970.

\bibitem{gelip} D. Galetti and A. F. R. de Toledo Piza, Physica {\bf
A214}, 207 (1995).

\bibitem{baires} M. C. Cambiaggio, G. G. Dussel and M. Saraceno,
Nucl. Phys. {\bf A415}, 70 (1984). 

\end{thebibliography}
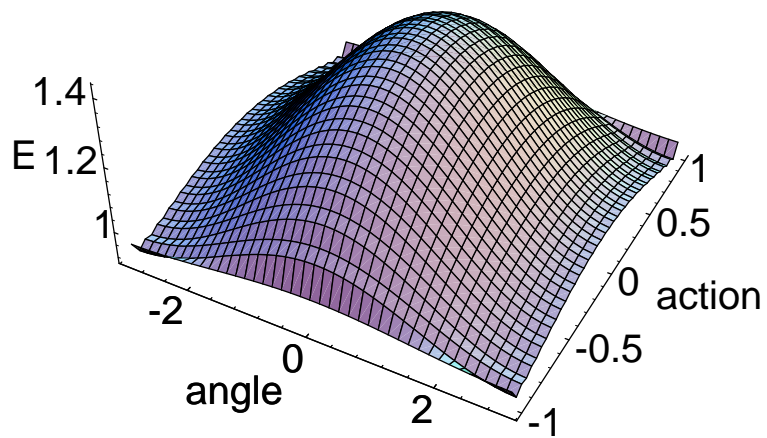
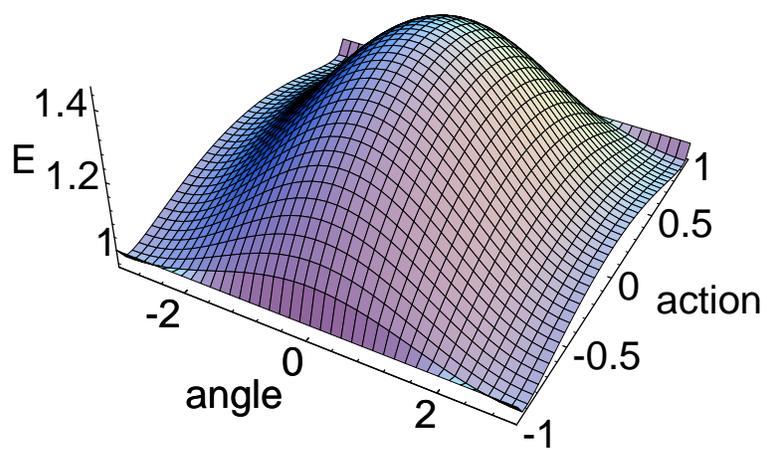
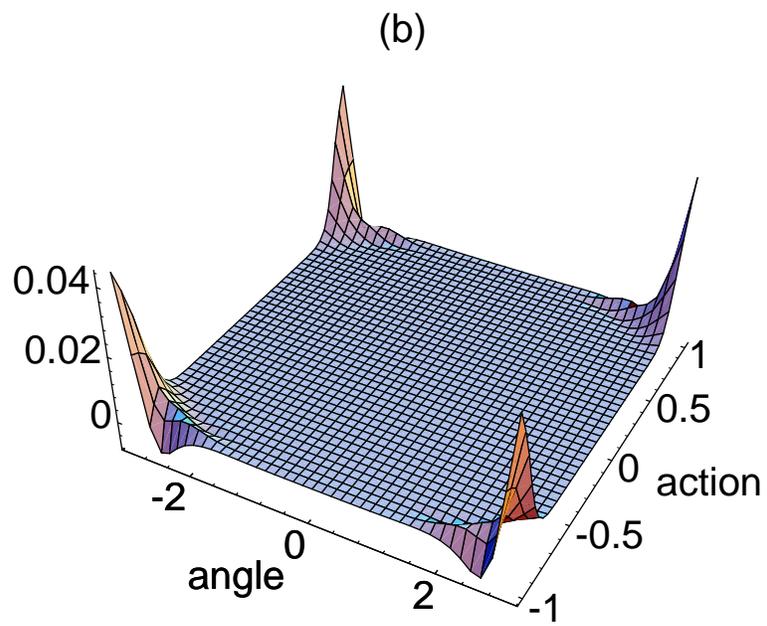
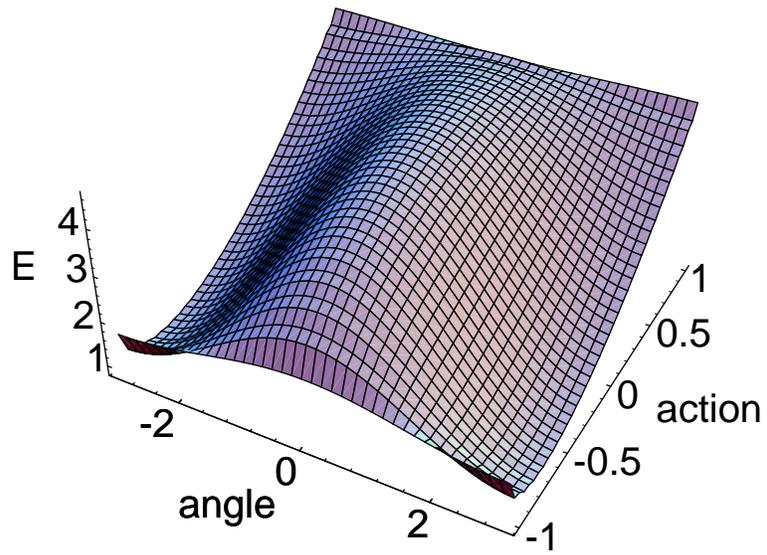
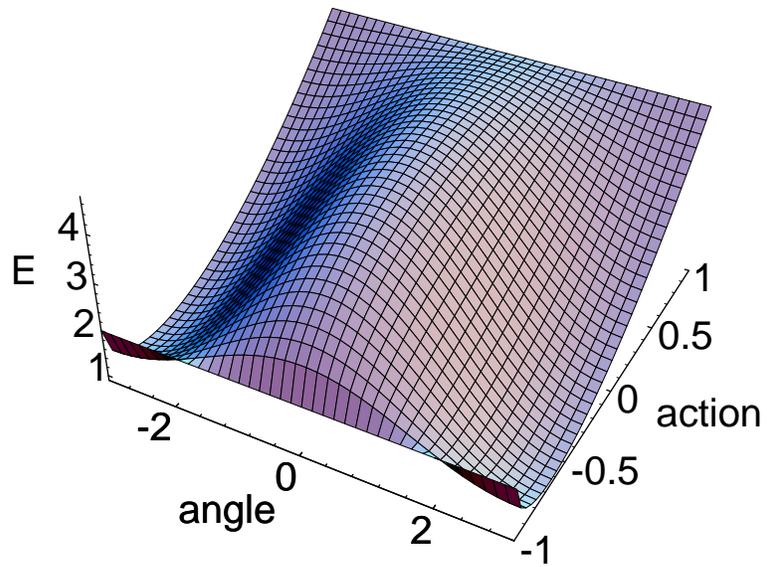
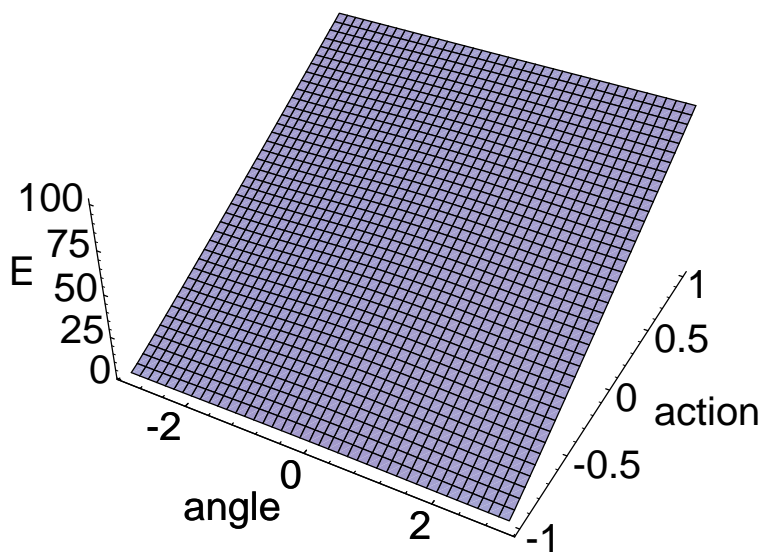
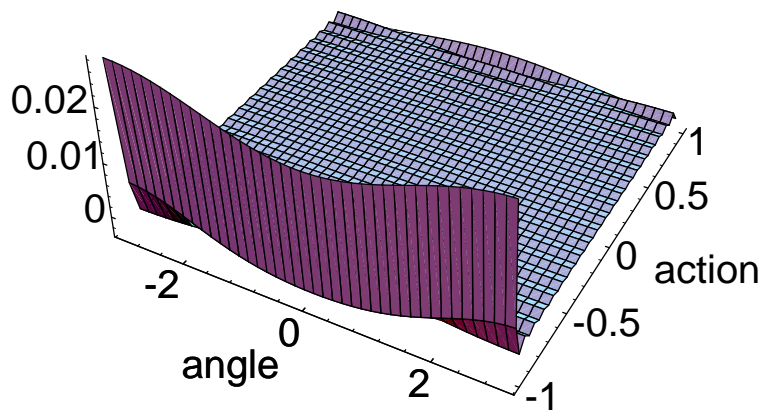
\end{document}